\documentstyle[12pt,bezier]{article}
\renewcommand{\a}{\alpha}

\renewcommand{\d}{\delta}         
\newcommand{\ve}{\varepsilon}

\newcommand{\ld}{\lambda}        \newcommand{\LD}{\Lambda}
\newcommand{\om}{\omega}         \newcommand{\OM}{\Omega}
              
\newcommand{\ro}{\rho}
           
\newcommand{\th}{\theta}         
           
\newcommand{\vf}{{\varphi}}

\newcommand{\cM}{{\cal M}}

\newcommand{\cO}{{\cal O}}

\newcommand{\hH}{{\widehat H}}

\newcommand{\Ht}{{\widehat t}}

\newcommand{\Hx}{{\widehat x}}

\newcommand{\ra}{\,\rightarrow\,}
\def\limar#1#2{\,\raise0.3ex\hbox{$\longrightarrow$\kern-1.5em\raise-1.1ex
\hbox{$\scriptstyle{#1\rightarrow #2}$}}\,}
\def\limarr#1#2{\,\raise0.3ex\hbox{$\longrightarrow$\kern-1.5em\raise-1.3ex
\hbox{$\scriptstyle{#1\rightarrow #2}$}}\,}
\def\limlar#1#2{\ \raise0.3ex
\hbox{$-\hspace{-0.5em}-\hspace{-0.5em}-\hspace{-0.5em}
\longrightarrow$\kern-2.7em\raise-1.1ex
\hbox{$\scriptstyle{#1\rightarrow #2}$}}\ \ }

\newcommand{\da}{{\dagger}}

\newcommand{\exx}[1]{\exp\left\{ {#1}\right\}}
\newcommand{\ord}[1]{\mbox{\boldmath{$\cO$}}\left({#1}\right)}

\newcommand{\Z}{Z\!\!\!Z}
\newcommand{\Ibb}[1]{ {\rm I\ifmmode\mkern
            -3.6mu\else\kern -.2em\fi#1}}
\newcommand{\ibb}[1]{\leavevmode\hbox{\kern.3em\vrule
     height 1.2ex depth -.3ex width .2pt\kern-.3em\rm#1}}

\newcommand{\R}{{\Ibb R}}

\newcommand{\rational}{{\kern .1em {\raise .47ex
\hbox{$\scripscriptstyle |$}}
    \kern -.35em {\rm Q}}}

\newcommand{\intf}{\int_{-\infty}^{\infty}\,}

\newcommand{\Ree}{{\cal R}\!e \,}
\newcommand{\Imm}{{\cal I}\!m \,}
\newcommand{\tr}{{\rm {Tr} \,}}
\newcommand{\er}{{\rm{e}}}
\renewcommand{\i}{{\rm{i}}}

\newcommand{\pa}{\partial}
\newcommand{\pad}[2]{{\frac{\partial #1}{\partial #2}}}

\def\<{\langle}
\def\>{\rangle}

\newcommand{\bra}[1]{\left\langle {#1}\right|}
\newcommand{\ket}[1]{\left| {#1}\right\rangle}

\newcommand{\be}{\begin{equation}}
\newcommand{\ee}{\end{equation}}
\newcommand{\bn}{\begin{eqnarray}}
\newcommand{\en}{\end{eqnarray}}
\newcommand{\bnn}{\begin{eqnarray*}}
\newcommand{\enn}{\end{eqnarray*}}
\newcommand{\ba}{\begin{array}}
\newcommand{\ea}{\end{array}}
\newcommand{\e}{\label}
\newcommand{\nbr}{\nonumber\\[3mm]}
\newcommand{\r}[1]{(\ref{#1})}

\newcommand{\qq}{\qquad}

\newcommand{\biz}{\begin{itemize}} 
\newcommand{\eiz}{\end{itemize}}
\newcommand{\ben}{\begin{enumerate}} 
\newcommand{\een}{\end{enumerate}}

\begin{document}
\begin{titlepage}
\
\vspace{0.5cm}
\begin{flushright}
July 19, 2000
\end{flushright}

\begin{center}
\vspace*{1.0cm}

{\Large \bf Space-Time Noncommutativity,\\ 
\vspace{5mm}
Discreteness of Time and Unitarity}

\vskip 1cm

{\large {\bf M. Chaichian}}, 
\ \ {\large{\bf 
A. Demichev}}\renewcommand{\thefootnote}{a}
\footnote{Permanent address: 
Nuclear Physics Institute, Moscow State University,
119899 Moscow, Russia},
\ \ {\large{\bf 
P. Pre\v{s}najder}}\renewcommand{\thefootnote}{b}
\footnote{Permanent address:
Department of Theoretical Physics, Comenius University,
Mlynsk\'{a} dolina, SK-84248 Bratislava, Slovakia}
\ \ and \ \ {\large{\bf A. Tureanu}}
\vskip 0.2cm

High Energy Physics Division, Department of Physics,\\
University of Helsinki\\
\ \ {\small\it and}\\
\ \ Helsinki Institute of Physics,\\
P.O. Box 9, FIN-00014 Helsinki, Finland

\end{center}

\setcounter{footnote}{0}

\vspace{0.5 cm}

\begin{abstract}
\normalsize

Violation of unitarity for noncommutative field theory on compact space-times
is considered. Although such theories are free of ultraviolet divergences, they
still violate unitarity while in a usual field theory such a violation occurs
when the theory is nonrenormalizable. The compactness of space-like coordinates
implies discreteness of the time variable which leads to appearance of
unphysical modes and violation of unitarity even in the absence of a 
star-product in the interaction terms. Thus, this conclusion holds also for
other quantum field theories with discrete time. Violation of causality, among
others, occurs also as the nonvanishing of the commutation relations between
observables at space-like distances with a typical scale of noncommutativity. 
While this feature allows for a possible violation of the spin-statistics
theorem, such a violation does not rescue the situation but makes the scale of
causality violation as the inverse of the mass appearing in the considered
model, i.e., even more severe. We also stress the role of smearing over the
noncommutative coordinates entering the field operator symbols. 

\end{abstract}

\vspace{0.5cm}


\vspace{0.5cm}


\end{titlepage}

\section{Introduction}

It is generally believed that the picture of space-time as a manifold $\cM$
should break down at very short distances of the order of the Planck length.
One possible approach to the description of physical phenomena at small
distances is based on noncommutative (NC) geometry of the space-time. It has
been  shown that the noncommutative geometry  naturally appears in string
theory with a nonzero  antisymmetric $B$-field \cite{SeibergW}.  Another
approach starting from the study of a relation  between measurements at very small
distances and black hole formations  has been developed in the works
\cite{DoplicherFR}.  The essence of the noncommutative geometry consists in
reformulating first the geometry in terms of commutative algebras of smooth
functions, and then generalizing them to their noncommutative analogs in terms
of  operators (or, more generally, to use a $C^*$-algebra) generated 
by noncommuting space and time coordinates: $[\Hx^\mu,\Hx^\nu]\neq0$. 

The Hilbert (Fock) space for a commutative and the corresponding NC field
theories are the same at the perturbative level. This is supported by the fact
that the quadratic part of the action is not affected by a star-product.
Moreover, this is the reason why there should be a map between any NC field
theory and its commutative limit: the degrees of freedom are the same. 

Noncommutative field theories with noncommutativity of only space coordinates
(while the time remains a usual commutative variable) do not change crucially
the standard quantum mechanical formalism (one can develop the usual
Hamiltonian dynamics, define the corresponding Schr\"odinger picture, etc.).
Of course, this kind of noncommutativity still essentially changes some
properties of the theory: in particular, it becomes nonlocal in the space-like
directions \cite{MinwallaRS,SeibergST}. But such basic properties of physical
models as causality and unitarity are satisfied. This can be traced back
\cite{MinwallaRS,SeibergST} to the fact that this theory describes low energy
excitations of a D-brane in the presence of a background magnetic field (see
\cite{SeibergW} and refs. therein). 

Field theories with space and {\it time} noncommutativity provide an
interesting opportunity to test the possible breakdown of the conventional
notion of time and the familiar framework of quantum mechanics at the Planck
scale. As it has been shown in the works \cite{SeibergST,GomisM,GomisKL},  in
the case of the model derived from string theory with a background electric
field and in the flat space-time, noncommutativity of the time coordinates  of
the corresponding Minkowski space  and the corresponding nonlocality in time
result in violation of both the causality and unitarity conditions. 

Thus, the question whether there exists some self-consistent theory with
noncommutative time coordinate is of great interest. The analysis in 
\cite{SeibergST,GomisM,GomisKL} shows that the violation of the basic
principles of causality and unitarity occurs at energies higher than the
inverse scale of the parameter of noncommutativity $\ld$,  \i.e. for
$E\gg\ld^{-1}$. Thus if some noncommutative theory implies an upper bound on
possible values of energy, one may hope that it is free of the problems with
the violation of the basic physical principles. In the paper \cite{CDP1}, we
had shown that space-time quantization on a two-dimensional cylinder leads to
the energy spectrum,  confined within the interval $E\in [0,\pi/\ld]$.
Therefore, it is natural  to study the question about unitarity and causality
for this case.  It is worth noticing that this restriction on energy provides
an improved ultraviolet behaviour of the field theory on the NC cylinder: even
planar diagrams in this case prove to be convergent (in contrast to the theory
in the flat NC Minkowski space).  

In fact, such a study has even wider interest. The point is that the
restriction on energy values appears as a consequence of discreteness of time
(in the representation where the time coordinate operator is diagonal). On the
other hand, attempts have been made to construct quantum field theories with
discrete time which is considered to be not only an intermediate 
regularization (as in the lattice field theories) but has fundamental physical
meaning \cite{Snyder} (for recent attempts see, e.g., the series of papers
\cite{JaroszkiewiczN} and refs. therein). The problem of unitarity has not been
investigated for this kind of models.

In this letter, we shall show that the situation with the violation of the
unitarity condition on the cylinder is even more severe than that in the case
of the flat space-time. More precisely, due to the discreteness of the time
evolution, the unitarity requirement is violated even by planar diagrams (which
do not carry a trace of the star-product). That means that the result is valid
for any theory with discrete time variable and not only for the field theory
with the space-time noncommutativity. 

The letter is arranged as follows. In section \ref{ftnc}, we present some  facts
about noncommutative cylinder and the corresponding $\Phi^4$-field theory,
necessary for further study. In section \ref{utdt}, we prove the  violation of
unitarity for planar diagrams in one-loop approximation.
Section \ref{dac} is devoted to conclusions and remarks. 

\section{Field theory on a noncommutative cylinder \e{ftnc}}
The points on a commutative cylinder $C$ can be specified by a
real parameter $t\in{\R}$ and two complex parameters $x_\pm
=\rho \er^{\pm\i\a}$. The fields possess the following expansion:
\be
\Phi (t,{\a} )=\sum_{k=-\infty}^{\infty} \intf
\frac{d\omega }{2\pi } {\tilde \Phi}_k (\omega )
\er^{\i k{\a}-\i\omega t}  \ .            \e{2.1}
\ee

In the noncommutative case \cite{CDP1} the parameters $t,x_\pm$ are replaced
by operators ${\hat t},{\hat x}_\pm$ satisfying the commutation
relations
\be
[{\hat t},{\hat x}_\pm ]=\pm\ld {\hat x}_\pm \ ,\qq 
[{\hat x}_+ ,{\hat x}_- ]\ =\ 0\ ,                            \e{2.2}
\ee
and the same constraint equation as in the commutative
case: ${\hat x}_+ {\hat x}_- =\rho^2$. The dimensionful (with the 
dimension of length) parameter $\ld$ is an analog of the 
tensor $\th$ in the case of the Heisenberg-like commutation relation in the
flat Minkowski space. 
However, in the present case, 
the actual parameter of the noncommutativity is the dimensionless 
parameter $\eta =\lambda /\rho$.

The operators ${\hat t},{\hat x}_\pm$ can be realized in the
auxiliary Hilbert space ${\cal H}=L^2 (S^1 ,d{\a})$ as follows:
\be
{\hat t}=-\lambda \partial_{\a} \ ,\qq {\hat x}_\pm = \rho
\er^{\pm\i k{\a}} \ .                                    \e{2.3}
\ee
We specify the self-adjoint extension of
$\partial_{\a}$ by postulating its system of eigenfunctions:
$\partial_{\a} f_k ({\a} )=\i k f_k ({\a} )$, 
$f_k ({\a})=\er^{\i k{\a}}$, 
$k\in{\Z}$. Thus, we are dealing with a 
unitary irreducible representation of the two-dimensional Euclidean group
$E(2)$ specified by the value of the Casimir operator 
${\hat x}_+{\hat x}_- =\rho^2$.

In analogy with the commutative case, we take the fields
to be operators in ${\cal H}=L^2 (S^1 ,d{\a})$
possessing the operator Fourier expansion:
\be
\Phi ({\hat t},{\hat {\a}})\ =\ \sum_{k=-\infty}^\infty 
\int_{-\pi/\lambda}^{+\pi /\lambda} \frac{d\omega }{2\pi } {\tilde \Phi}_k
(\omega ) \er^{\i k{\hat{\a}}-\i\omega{\hat t}}\ .       \e{2.4}
\ee
For simplicity, we shall consider a real scalar field theory which corresponds
to the condition $\Phi^\da ({\hat t},{\hat {\a}})=\Phi ({\hat t},{\hat
{\a}})$. It is important that since the spectrum of ${\hat t}$ is discrete:
$t=\lambda n$, $n\in{\Z}$, the integration over $d\omega$ goes only over a
finite interval $(-\pi /\lambda ,+\pi /\lambda )$.  We point out that the
operator Fourier expansion \r{2.4} is invertible:
\be
{\tilde \Phi}_k (\omega )=\frac{1}{2\pi }{\tr}\Big[
\er^{-\i k{\hat{\a}}+\i\omega{\hat t}} \Phi ({\hat t},{\hat
{\a}})\Big]\ .                                         \e{2.5}       
\ee
This follows straightforwardly from the formula
\be
\frac{1}{2\pi }{\tr}[ \er^{-\i k'{\hat{\a}}+\i\omega'{\hat t}}
\er^{\i k{\hat{\a}}-\i\omega{\hat t}} ]=\delta_{k'k} \delta^{(S)}
(\lambda\omega' -\lambda\omega )\ ,                        \e{2.6}
\ee
where $\delta^{(S)}(\vf)$ denotes the $\delta$-function on a circle.
The inverse {\it usual} Fourier transform of ${\tilde \Phi}_k (\omega)$ yields
an analog of the Weyl symbol $\Phi(n\ld,{\a})$ on the cylinder:
\be
\Phi (n\ld,{\a})\ =\ \sum_{k=-\infty}^\infty 
\int_{-\pi/\lambda}^{+\pi /\lambda} \frac{d\omega }{2\pi } {\tilde \Phi}_k
(\omega ) \er^{\i k{{\a}}-\i\ld\omega n}\ .       \e{2.7}
\ee
Notice that since $\Phi(n\ld,{\a})$ is not a function on the whole commutative
cylinder, but takes values only at discrete points of the time variable, this is
not the canonical Weyl symbol. The latter can be constructed if one considers
all possible self-adjoint extensions of the operator $\pa_{\a}$ on a circle.
Since this is not important for our consideration, we drop further discussion
of this possibility.

The star-product for the fields $\Phi(n\ld,{\a})$ has the form which is very
close to that appearing in the flat space-time:
\be
\Phi_1(n\ld,{\a})\star \Phi_2(n\ld,{\a})=
\er^{\frac{\i\ld}{2}\left(\pad{\ }{t_1}\pad{\ }{\vf_2}
-\pad{\ }{t_2}\pad{\ }{\vf_1}\right)}
\Phi_1(n\ld+t_1,{\a}+\vf_1) 
\Phi_2(n\ld+t_2,\a+\vf_2)\Bigg|_{\stackrel{\scriptstyle t_1=t_2=0}
{\scriptstyle\vf_1=\vf_2=0}}\ ,\e{2.8}
\ee
where $t_1,t_2,\vf_1,\vf_2$ are auxiliary continuous variables.

On the commutative cylinder, the d'Alembertian can be
expressed through the Poisson brackets \cite{CDP1}:
\be
\Box\Phi=\{t,\{t,\Phi\}\}+\ro^{-2}\{x_+,\{x_-,\Phi\}\}\ ,   \e{2.9}
\ee
where $\{F,G\} =\frac{\partial F}{\partial\varphi}
\frac{\partial G}{\partial t}-\frac{\partial F}{\partial t}
\frac{\partial G}{\partial\varphi}$. We generalize it to the
noncommutaive case by replacing the Poisson brackets by
commutators: $\{ .,.\}\to\frac{1}{\i\lambda}[.,.]$. This gives the
free action on the noncommutative cylinder in the form
\bn
S_0^{(NC)}[\hat\Phi]&=&\pi\eta\tr
\Big\{-\frac{1}{\ld^2}[\Hx_+,\hat\Phi][\Hx_-,\hat\Phi]
+\frac{1}{\ld^2}[\Ht,\hat\Phi]^2-{\mu}^2\hat\Phi^2\Big\}\nbr
&=&\frac{\eta}{2}\sum_{n=-\infty}^\infty\int_{-\pi}^\pi\,d{\a}\,
\Big[\big(\d\Phi(n,{\a})\big)^2-
\left(\pad{\Phi(n,{\a})}{{\a}}\right)^2-{\mu}^2\Big]\ .  \e{2.10}
\en
Here 
$$
\d\Phi(n,{\a})=\frac{1}{\eta}\Big[\Phi(n+1,{\a})-\Phi(n,{\a})\Big]
$$
(we have simplified the notation for the field: 
$\Phi(n\ld,{\a})\ra\Phi(n,{\a})$),
and ${\mu}$ is the dimensionless parameter related to the mass: ${\mu}=\ro m$.
As usual for the Weyl symbol, the star-product disappears from the trace 
for a product of any two operators. In the case of a field theory in a 
flat space, this leads to the free action which formally looks as the one 
on commutative space. In the case of cylinder, we have the trace 
of noncommutativity even in the free action: it reveals itself in 
discrete-time derivatives. We stress that this is an intrinsic property 
of field theories on noncommutative manifolds with {\it compact} space-like 
dimensions and appears in any formalism and for any operator symbols.

The $\Phi^4$-interaction term contains, in general, the star-product:
\bn
S_{int}^{(NC)}&=&\frac{g}{4!}\,2\pi\tr\Big\{\Phi^4(\Ht,\hat{\a})\Big\}\nbr
&=&\frac{g\eta}{4!}\sum_{n=-\infty}^\infty\int_{-\pi}^\pi\,d{\a}\,
\big(\Phi(n,{\a})\star\Phi(n,{\a})\big)^2\ .                  \e{2.11}
\en
In the momentum representation, this star-product results 
in the appearance of the factors 
$\cos[\ld(\om k'-\om' k)]$ (here $\om,\ \om',\ k,\ k'$ are 
the energies and momenta entering the vertex). 
These factors grow both in the upper and lower half-planes of 
the complex-energy plane and prevent us from the use of the standard 
Cutkosky cutting rules and, eventually, lead to the violation of unitarity. 
Although in the case of the cylinder, we have to consider only a strip 
$\Ree\om\in[-\pi/\ld,\pi/\ld]$ in the complex-energy plane, 
the consideration proves to be essentially the same as in the case 
of flat space-time \cite{GomisM} and we do not repeat it.

For a possible attempt to rescue the theory, 
one may try to modify the interaction
term. One possibility is to define the action through a specific ordering
prescription for the noncommuting operators $\Ht$ and $\hat{\a}$ (a situation
not obtainable from the known string theories). In
particular, a $t{\a}$-``normal'' ordering 
(i.e., the requirement that in the operator
expression for the action all operators $\Ht$ be posed to the left of 
all operators $\hat{\a}$) 
leads to disappearance of star product in the interaction term \cite{CDP1}:
\be
S_{int}^{(NC,t{\a})}
=\frac{g\eta}{4!}\sum_{n=-\infty}^\infty\int_{-\pi}^\pi\,d{\a}\,
\Phi^4(n,{\a})\ .                  \e{2.12}
\ee
In a flat space-time, such version of noncommutative field theory exactly
coincides with the usual commutative QFT (except that now one deals with
operator symbols instead of usual fields, so that interpretation of events in
space and time requires additional smearing, 
while all calculations and results in
the momentum space remain the same as in the usual QFT). 
On the contrary, in the
case of the cylinder, even after the ordering, we still have the trace of
the noncommutativity, namely, the discreteness of the time variable.  Thus, it
is interesting to verify (see next section) 
whether such variant of the noncommutative field
theory preserves unitarity. Another motivation for this study is
the persistent attempts to construct quantum field theories with improved
ultraviolet behaviour starting from the postulate of discreteness of time
\cite{JaroszkiewiczN}.

\section{Unitarity in theories with discrete time\e{utdt}}
The free field equation of motion derived from the action 
\r{2.10} reads as follows:
\be
\Big(\bar\d\d-\pa_{\a}\pa_{\a}+{\mu}^2\Big)\Phi(n,{\a})=0\ ,       \e{2.13}
\ee
(here $\bar\d f(n)\equiv[f(n)-f(n-1)]/\eta$) and the corresponding propagator
has the form
\be
D_0^{(NC)}(\om,k)=\frac{1}{\OM^2(\om)-k^2/\ro^2-m^2+\i\ve}\ ,       \e{2.14}
\ee
where 
\be
\OM=\frac{2}{\ld}\sin\left(\frac{\ld\om}{2}\right)\ .          \e{2.15}
\ee
The modes which satisfy the condition $k^2\leq \LD^2\equiv 4/\eta^2-{\mu}^2$
correspond to the usual oscillating solutions of the equation \r{2.13} and
resemble the solutions in the continuous-time physics. On the contrary, the
modes with $k^2> \LD^2\equiv4/\eta^2-{\mu}^2$ correspond 
to growing or decreasing 
in time solutions and, as we shall show soon, are unphysical. Correspondingly,
the propagator has two types of singular points: 
\biz
\item[--] the oscillating modes with $k^2\leq\LD^2$ produce poles in the
complex-energy plane at $\pm\om_k\mp\i\ve$, where $\om_k>0$ is defined by the
equality: 
$$
\sin^2\left(\frac{\ld\om_k}{2}\right)=\frac{\eta^2}{4}\big(k^2+{\mu}^2\big)\ ;
$$
\item[--] the modes with $k^2 >\LD^2$ produce poles at 
$\om_k=\pi/\ld\pm\i S_k$, where $S_k>0$ is defined by the equality 
\be
\cosh(\ld S_k/2)=\frac{\eta^2}{2}(k^2+{\mu}^2)-1\ .                \e{2.15a}
\ee
\eiz

In order to realize the physical meaning of the two types of the modes, we use 
the method of the transfer matrix (see, e.g. \cite{MontvayM}). The transfer
matrix $T_k$ for a given mode
$\Phi_k(n)=(2\pi)^{-1}\int\,d\a\,\Phi(n,\a)\exx{-\i k\a}$ in the discrete-time
field theory under consideration has the form:
$$
T_k=\exx{\i\left[\frac{\big(\Phi_k(n+1)-\Phi_k(n)\big)^2}{\eta}
-\frac{\eta}{2}(k^2+{\mu}^2)\Big(\Phi_k^2(n+1)+\Phi_k^2(n)\Big)\right]}\ .
$$
Then the calculation of the corresponding Hamiltonian, defined by
$\hH=-\i/\ld\ln T$, shows that while for the oscillating modes we obtain a
harmonic oscillator Hamiltonian with the frequency $W$ defined by the relation
$\sin(W\eta/2)=\eta\sqrt{k^2+{\mu}^2}/2$,  
the modes with $k^2 >4/\eta^2-{\mu}^2$ correspond to a Hamiltonian 
which is not a positive definite (bounded from below) operator. 
Thus these modes are unphysical ones and we have to study
unitarity within the subspace of the oscillating modes. In other words, we have
to check that the unphysical states decouple from the physical ones 
similarly to the ghost fields in the gauge field theory or to unstable 
states \cite{Veltman}.

We shall check the unitarity condition, i.e.
\be
2\Imm M_{ab}=\sum_cM_{ac}M_{cb}\ ,               \e{2.15b}
\ee
for the on-shell transition matrix elements $M_{ab}$ between states $a$ and $b$
in second order of the perturbation theory for the interaction of the
form \r{2.12} (i.e., for planar diagrams in the case of the standard
noncommutative field theory or for a theory with the $t{\a}$-ordering defined 
above, or for a theory which simply starts from postulating discreteness 
of time).

One can easily check that at the tree level the unitarity condition in
the physical sector is indeed satisfied:
\par
\unitlength=1mm
\linethickness{0.4pt}
\begin{picture}(59.67,18.00)(-20.00,0.00)
\put(11.67,10.33){\circle*{0.67}}
\put(11.67,10.33){\line(-1,1){4.67}}
\put(11.67,10.33){\line(-1,0){4.67}}
\put(11.67,10.33){\line(-1,-1){4.33}}
\put(11.67,10.33){\line(1,0){9.00}}
\put(20.67,10.33){\line(1,1){4.33}}
\put(20.67,10.33){\line(1,-1){4.00}}
\put(20.67,10.33){\line(1,0){4.00}}
\put(20.67,10.33){\circle*{0.67}}
\put(4.33,10.33){\makebox(0,0)[rc]{$2\Imm$}}
\put(31.67,10.33){\makebox(0,0)[cc]{$=$}}
\put(45.67,8.33){\makebox(0,0)[cc]{${\displaystyle\sum_{k<4/\eta^2-{\mu}^2}}$}}
\put(55.33,4.33){\line(0,1){11.67}}
\put(62.33,10.33){\circle*{0.67}}
\put(62.33,10.33){\line(-1,1){4.67}}
\put(62.33,10.33){\line(-1,0){4.67}}
\put(62.33,10.33){\line(-1,-1){4.33}}
\put(62.67,10.33){\line(1,0){4.00}}
\put(65.33,8.33){\makebox(0,0)[cc]{$k$}}
\put(68.67,4.33){\line(0,1){11.67}}
\put(69.67,15.67){\makebox(0,0)[lc]{{\large 2}}}
\end{picture}
\par\noindent 
Next, we consider the $s$-channel 1-loop Feynman diagram
\par
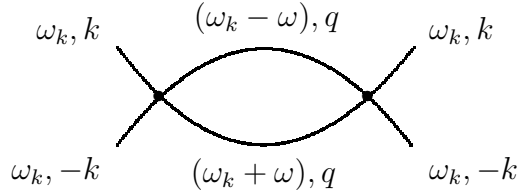
\unitlength=1.3mm
\linethickness{0.4pt}
\begin{picture}(38.33,26.67)(-20.00,0.00)
{\thicklines 
\bezier{196}(6.33,7.00)(21.33,26.67)(36.33,7.00)
\bezier{200}(6.33,17.00)(21.00,-3.00)(36.67,17.00)}
\put(10.67,12.00){\circle*{1.00}}
\put(32.00,12.00){\circle*{1.00}}
\put(4.67,18.67){\makebox(0,0)[rc]{$\om_k,k$}}
\put(4.67,4.33){\makebox(0,0)[rc]{$\om_k,-k$}}
\put(38.33,18.67){\makebox(0,0)[lc]{$\om_k,k$}}
\put(38.33,4.33){\makebox(0,0)[lc]{$\om_k,-k$}}
\put(21.67,20.00){\makebox(0,0)[cc]{$(\om_k-\om),q$}}
\put(21.33,4.00){\makebox(0,0)[cc]{$(\om_k+\om),q$}}
\end{picture}
\par\noindent
in the center-of-mass frame (one can easily check that the corresponding $t$-
and  $u$-channel diagrams have no branch cut singularities above the threshold).
The corresponding contribution to the matrix element reads
\be
\i M=\frac{g^2}{2}\sum_{q=-\infty}^\infty\int_{-\pi/\ld}^{\pi/\ld}\,d\om\,
D_0^{(NC)}(\om_k+\om,q)D_0^{(NC)}(\om_k-\om,q)\ .     \e{2.16}
\ee
The calculation of the imaginary part of the amplitude can be carried out by
closing the contour of integration in the complex-energy plane downward as it
is shown in figure~\ref{F:1}.
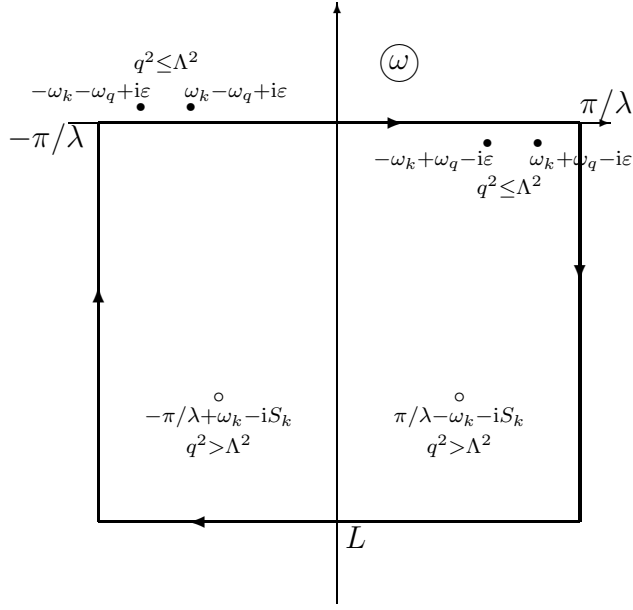
\begin{figure}
\centering
\unitlength=0.8mm
\linethickness{0.4pt}
\begin{picture}(93.33,103.33)
\put(48.00,3.33){\vector(0,1){100.00}}
\put(3.33,83.33){\vector(1,0){90.00}}
\put(58.33,93.33){\circle{6.00}}
\put(58.33,93.33){\makebox(0,0)[cc]{$\om$}}
\put(73.00,80.00){\circle*{1.33}}
\put(81.33,80.00){\circle*{1.33}}
\put(15.33,86.00){\circle*{1.33}}
\put(23.66,86.00){\circle*{1.33}}
\put(68.33,37.66){\circle{1.50}}
\put(28.33,37.66){\circle{1.50}}
\thicklines
\put(8.33,83.33){\line(0,-1){66.33}}
\put(8.33,17.00){\line(1,0){80.00}}
\put(88.33,17.00){\line(0,1){66.33}}
\put(88.33,83.33){\line(-1,0){80.00}}
\put(53.66,83.33){\vector(1,0){5.33}}
\put(88.33,64.00){\vector(0,-1){7.00}}
\put(36.33,17.00){\line(-1,0){10.33}}
\put(33.66,17.00){\vector(-1,0){10.33}}
\put(8.33,45.00){\vector(0,1){11.00}}
\put(6.00,80.66){\makebox(0,0)[rc]{$-\pi/\ld$}}
\put(88.33,86.33){\makebox(0,0)[lc]{$\pi/\ld$}}
\put(49.33,14.33){\makebox(0,0)[lc]{$L$}}
\put(74.00,77.00){\makebox(0,0)[rc]{${\scriptstyle -\om_k+\om_q-\i\ve}$}}
\put(80.00,77.00){\makebox(0,0)[lc]{${\scriptstyle \om_k+\om_q-\i\ve}$}}
\put(17.00,88.33){\makebox(0,0)[rc]{${\scriptstyle -\om_k-\om_q+\i\ve}$}}
\put(22.66,88.33){\makebox(0,0)[lc]{${\scriptstyle \om_k-\om_q+\i\ve}$}}
\put(68.33,32.00){\makebox(0,0)[cc]
{\shortstack{${\scriptstyle \pi/\ld-\om_k-\i S_k}$
\\${\scriptstyle q^2>\LD^2}$}}}
\put(28.33,32.00){\makebox(0,0)[cc]
{\shortstack{${\scriptstyle -\pi/\ld+\om_k-\i S_k}$\\
${\scriptstyle q^2>\LD^2}$}}}
\put(76.66,72.66){\makebox(0,0)[cc]{${\scriptstyle q^2\leq\LD^2}$}}
\put(19.66,93.33){\makebox(0,0)[cc]{${\scriptstyle q^2\leq\LD^2}$}}
\end{picture}
\caption{The singularities for the two type of modes and the contour of
integration for the calculation of the imaginary part of the amplitude;
$\om_q=(2/\ld)\mbox{arcsin}[(\eta/2)\protect\sqrt{q^2+\mu^2}]$.\e{F:1}}
\end{figure}

In figure~\ref{F:1}, the filled circles denote the usual Feynman-like poles at
the points $\om=\pm\om_k+(2/\ld){\rm arcsin}[(\eta/2)\sqrt{k^2+{\mu}^2}]-\i\ve$
(in the lower half-plane) and at  $\om=\pm\om_k-(2/\ld){\rm
arcsin}[(\eta/2)\sqrt{k^2+{\mu}^2}]+\i\ve$ (in the upper half-plane), appearing
for the oscillating modes with $q^2\leq \LD^2$.    The small empty circles
denote position of the singularities for the unphysical modes with $q^2>\LD^2$
at $\om=(\pm\pi/\ld\mp\om_k)-\i S_k$. Here $S_k>0$ is the solution of the
equation \r{2.15a} (there exist symmetrical singularities in the upper
half-plane but they are not important for us). The closing of the contour is
possible due to the facts that the contributions from its vertical parts cancel
each other due to the periodicity in the energy variable, while the lower
horizontal part gives a vanishing contribution when the distance $L$ to the
real axis goes to infinity. The latter is true only for the interaction vertex
{\it without} the star-product cosine factors (i.e. for planar diagrams, or for
theories with the $t\a$-``normal'' ordering defined above or simply for a
theory with  discrete time).   

We separate the sum in \r{2.16} into two parts: $\sum_{q\in\Z}= \sum_{|q|\leq
\LD}+\sum_{|q|>\LD}$ and, first, we consider the part with the oscillation
modes $|q|\leq \LD$. Then, proceeding in the usual way \cite{Cutkosky} (see
also, e.g. \cite{EdenLOP} and refs therein) and taking the residues of the
corresponding poles, one can show that {\it this} part of the sum already gives
the contribution which saturates the unitarity condition in the physical sector
of the oscillating modes:
\par
\unitlength=1.00mm
\linethickness{0.4pt}
\begin{picture}(89.00,28.67)(-20.00,0.00)
\bezier{196}(11.66,9.00)(26.66,28.67)(41.66,9.00)
\bezier{200}(11.66,19.00)(26.33,-1.00)(42.00,19.00)
\put(16.00,14.00){\circle*{0.67}}
\put(37.00,14.00){\circle*{0.67}}
{\thicklines
\bezier{88}(9.00,24.00)(5.00,14.67)(9.00,4.00)
\bezier{88}(46.00,24.00)(51.00,14.00)(46.00,4.00)}
\put(2.67,13.67){\makebox(0,0)[rc]{$2\,\Imm$}}
\put(31.00,7.33){\makebox(0,0)[lc]{${\sum_{|q|\leq \LD}}$}}
\put(26.00,5.00){\line(0,1){2.00}}
\put(26.00,8.00){\line(0,1){2.00}}
\put(26.00,11.00){\line(0,1){2.00}}
\put(26.00,14.00){\line(0,1){2.00}}
\put(26.00,17.00){\line(0,1){2.00}}
\put(26.00,20.00){\line(0,1){2.00}}
\put(54.33,14.00){\makebox(0,0)[cc]{$=$}}
\put(61.33,12.33){\makebox(0,0)[cc]{${\displaystyle\sum_{|p_{1,2}|\leq \LD}}$}}
\put(80.00,14.67){\line(-1,1){6.67}}
\put(80.00,14.67){\line(1,1){6.33}}
\put(80.00,14.67){\line(-1,-1){6.00}}
\put(80.00,14.67){\line(1,-1){5.67}}
\put(80.00,14.67){\circle*{0.67}}
\put(87.67,19.33){\makebox(0,0)[cc]{${\scriptstyle p_1}$}}
\put(87.67,9.00){\makebox(0,0)[cc]{${\scriptstyle p_2}$}}
{\thicklines
\put(92.67,3.67){\line(0,1){20.00}}
\put(69.67,4.33){\line(0,1){20.00}}}
\put(94.00,23.33){\makebox(0,0)[lc]{{\large 2}}}
\end{picture}
\par

Unfortunately, the part of the sum corresponding to the unphysical modes also
gives a contribution to the imaginary part of the amplitudes due to the poles
indicated in the figure~\ref{F:1} by the empty circles. In general, this
non-zero contribution looks rather cumbersome, but for the particular value of
the external energy, namely for $\om_k=\pi/(2\ld)$, it becomes quite simple:
$$
\Big[2\Imm M\Big]^{(unph)}\Big|_{\om_k=\pi/(2\ld)}
=\frac{g^2}{4(2\pi)^2}\sum_{|q|>\LD}\big(q^2+{\mu}^2\big)^{3/2}
\big(q^2+{\mu}^2-4/\eta^2\big)^{-1/2}\ .
$$
The proof of the unitarity violation for theories with flat space-like
dimensions (e.g., for the one proposed in \cite{JaroszkiewiczN}) goes
essentially in the same way (with the only distinction that the sums over
momentum modes is substituted by the corresponding integrals).  Thus, the
theories with a discrete time variable do not satisfy the unitarity condition.

\section{Conclusions and remarks\e{dac}}

We have shown that the transition to noncommutative spaces with compact
space dimensions does not help in restoring unitarity in the theories
with space and {\it time} noncommutativity. We also have proved a more general
statement that any theory with discrete time variable meets the same problem. 

It is clear from the previous section, 
that in the absence of the cosine factors in vertices, coming from
the star-product, the origin of the nonunitarity in theories with discrete time
is the appearance of the unphysical nonoscillating modes. Notice that if one
takes a specific value of the parameter of noncommutativity, namely
$\eta=2\pi/N$ ($N$ is a positive integer), the basic operator exponentials in
\r{2.4} satisfy the commutation relation
$$
\er^{\i \hat t}\er^{\i\hat{\a}}=
\er^{\i 2\pi/N}\er^{\i\hat{\a}}\er^{\i \hat t}\ ,
$$
and possess finite-dimensional representations \cite{Schwinger}. This implies
that for a small mass $m\sim\ord{N^{-2}}$ appearing in \r{2.10}, 
we can get rid of the unphysical modes. However, the
choice of the noncommutativity parameter as indicated above, means, actually,
transition to the quantum torus \cite{ConnesR}, i.e., to a manifold with
closed (compact) time-like curves. As is well-known \cite{MorrisTYH}, 
theories on such manifolds, even in
the commutative case, have their own problems with causality and formulation of
the unitarity condition. Therefore, we do not pursue this possibility further
here.

Theories with space-time noncommutativity suffer also from the violation of 
causality. In the work \cite{SeibergST}, this fact was demonstrated on the 
example of the scattering of wave-packets. Another possibility to see the 
violation of (micro)causality is to calculate the matrix elements of equal-time
commutators of some observables in this theory.  We note that in physical
applications one has finally to smear over the noncommutative coordinates in
field operator symbols since the symbols themselves  do not reflect the values
of the operator coordinates \cite{CDP1}: 
$\overline{\Phi(x)}\equiv\bra{x}\hat\Phi(\Hx)\ket{x}$, where $\ket{x}$ is, for
instance, a (maximally localized) coherent state. This smearing would make a
difference in the interpretation of violation of, e.g., (micro)causality if the
violation would be occurring only at the scales of the order of $\ld$ and not
growing with the energy.

In particular, for
quadratic observables we have\footnote{Notice that all the vacuum expectation
values (vacuum-vacuum matrix elements) of the commutators between 
observables at space-like distances identically vanish for the NC 
field theories exactly like in the commutative case.}:
\be
\bra{0}[\overline{\Phi^2(x)},\, \overline{\Phi^2(y)}
\ket{\vec p,\vec k}\Big|_{x^0=y^0}\approx
\er^{-(\vec x-\vec y)^2/(4\ld^2)}\ .                       \e{concl1}
\ee
The asymptotic behaviour \r{concl1} has been derived for the case when 
the distance between two points is large: $|\vec x-\vec y|\gg\ld$, and the
momenta $\vec p,\vec k$ are not too high (the result looks similar for both 
cases of a flat space-time and the cylinder, if the distance is understood
accordingly). For large values of momenta $\vec p$ and $\vec k$ of a
two-particle state $\ket{\vec p,\vec k}$, however, 
the exponential damping \r{concl1}
does not occur anymore. Actually, this violation of the causality 
(as well as that
observed in \cite{SeibergST}) can be interpreted as impossibility of precise
simultaneous measurement of space and time coordinates, in accordance with the
original idea presented in \cite{DoplicherFR}.
We also mention that in NC theory with
$t\a$-``normal''(``time-space'') ordering prescription all the commutators
between observables would vanish at space-like distances. 

Another interesting question concerning the NC field theories is the problem of
causality and the spin-statistics theorem \cite{Pauli}  (see also, e.g.,
\cite{SW}). As it is well-known, in the usual commutative quantum field theory
the requirement of vanishing of commutators for physical observables at
space-like distances (i.e., causality) leads uniquely to the spin-statistics
theorem. Since in NC field theory such commutation relations are not equal to
zero as explained above, one has, in principle, no more the same arguments for
the derivation of the spin-statistics relation and thus the modification of the
latter is not excluded\footnote{It is interesting  that the $CPT$-theorem
remains valid in NC field theories \cite{SJabbari}, but it is known that the
$CPT$-theorem requires weaker assumptions than the spin-statistics one does.}. 
We have studied several most natural modifications of the usual spin-statistics
(i.e., modifications of the commutation relations for creation and annihilation
operators) and found out that they are not only unable to help in the
restoration of the causality (cf. \r{concl1}) but instead they lead to
commutation relations which are nonvanishing as in \r{concl1} but with a scale
of the mass of the field $m$ instead of $1/\ld$ as in \r{concl1}, which is even
a more severe violation of causality. This violation of (micro)causality is of
exactly the same form which occurs in the usual commutative field theories when
one modifies the spin-statistics relation.   

\vspace{5mm}

{\bf Acknowledgments} 
We are much grateful to M. Sheikh-Jabbari for discussions and useful comments.
The financial support of the Academy of Finland under the Project No. 163394 
is greatly acknowledged. 
A.D.'s work was partially supported by RFBR-00-02-17679 grant
and P.P.'s work by VEGA project 1/7069/20.


\begin{thebibliography}{99}
\bibitem{SeibergW} N. Seiberg and E. Witten, {\it JHEP} 
{\bf  9909: 032} (1999); {\it hep-th/9908142}.
\bibitem{DoplicherFR}S.~Doplicher, K.~Fredenhagen and J.~E.~Roberts,
 {\it Phys. Lett.} {\bf B331} (1994) 39;\\
S.~Doplicher, K.~Fredenhagen and J.~E.~Roberts,
 {\it Comm. Math. Phys.} {\bf 172} (1995) 187.
\bibitem{MinwallaRS} S.Minwalla, 
M. Van Raamsdonk and N. Seiberg, {\it Noncommutative
Perturbative Dynamics}, PUPT-1905, IASSNS-HEP-99-112; {\it hep-th/9912072}. 
\bibitem{SeibergST} N. Seiberg, L. Susskind and N. Toumbas, 
{\it JHEP} {\bf 0006:044}; 
{\it hep-th/0005015}. 
\bibitem{GomisM} J. Gomis and T. Mehen, {\it Space-Time Noncommutative Field
Theories and Unitarity}, CALT-68-2272; {\it hep-th/0005129}
\bibitem{GomisKL} J. Gomis, K Kamimura and J. Llosa, 
{\it Hamiltonian Formalism for Space-Time Noncommutative Theories}, 
CERN-TH/2000-186; {\it hep-th/0006235}.
\bibitem{CDP1} M. Chaichian, A. Demichev and P. Pre\v{s}najder, 
Nucl. Phys. B567 (2000) 360.
\bibitem{Snyder} H. Snyder, {\it Phys. Rev.} {\bf 71} (1947) 38;\\
T. D. Lee, {\it Phys. Lett.} {\bf 122B} (1983) 217;\\
G. {}'t Hooft, {\it Class. Quant. Grav.} {\bf 10} (1993) 1023; 
{\it Class. Quant. Grav.} {\bf 10} (1993) Suppl. 79.
\bibitem{JaroszkiewiczN} 
G. Jaroszkiewicz and K. Norton, {\it J. Phys.} {\bf A30} (1997) 3115; 
{\it ibid.} {\bf A30} (1997) 3145; {\it ibid.} {\bf A31} (1998) 977; 
{\it ibid.} {\bf A31} (1998) 1001; {\it hep-th/9804165}.
\bibitem{MontvayM} J. B. Kogut {\it Rev. Mod. Phys.} {\bf 51} (1979) 659.
\bibitem{Veltman} M. Veltman, {\it Physica} {\bf 29} (1963) 186.
\bibitem{Cutkosky} R. E. Cutkosky, {\it J. Math. Phys.} {\bf 1} (1960) 429.
\bibitem{EdenLOP} R. J. Eden, P.V. Landshoff, D. I. Olive 
and J. C. Polkinghorn, 
{\it The Analytic S-Matrix} (Cambridge University Press, Cambridge, 1966).
\bibitem{Schwinger} J. Schwinger, 
{\it Proc. Nat. Acad. Sci.} {\bf 46} (1960) 570, 893; 
{\it ibid.} {\bf 47} (1961) 1075.
\bibitem{ConnesR} A. Connes, M. Rieffel, 
{\it Contemp. Math.} {\bf 62} (1987) 237;\\
M. Rieffel {\it Contemp. Math.} {\bf 105} (1991) 191. 
\bibitem{MorrisTYH} M. S. Morris, K. S. Thorne and U. Yurtsever, {\it Phys. Rev.
Lett.} {\bf 61} (1988) 1446;\\
S. W. Hawking, {\it Phys. Rev.} {\bf D46} (1992) 603.
\bibitem{Pauli} W. Pauli, {\it Phys. Rev.} {\bf 58} (1940) 716.
\bibitem{SW} R. F. Streater and A. S. Wightman, 
{\it PCT, Spin and Statistics, and All That} (Benjamin, New York, 1964).
\bibitem{SJabbari} M. M. Sheikh-Jabbari, {\it Discrete Symmetries (C,P,T) 
in Noncommutative Field Theories}, ICPT-2000-04; {\it hep-th/0001167}. 
\end{thebibliography}
\end{document}